\documentclass[aps,prl,twocolumn,showpacs,preprintnumbers,amsmath,amssymb]{revtex4}
\usepackage{graphicx}

\begin{document} 

\title{Absorbing Phase Transitions of Branching-Annihilating
Random Walks}

\author{Julien Kockelkoren and Hugues Chat\'e}

\affiliation{CEA -- Service de Physique de l'\'Etat Condens\'e, Centre d'\'Etudes de Saclay, 91191 Gif-sur-Yvette, France}

\date{\today}

\begin{abstract}
The phase transitions to absorbing states of the branching-annihilating
reaction-diffusion processes $mA\to (m+k)A$, $nA \to (n-l)A$ are studied
systematically in one space dimension within a new family of models. 
Four universality classes of non-trivial critical behavior are found. 
This provides, in particular, the first evidence of universal 
scaling laws for pair and triplet processes.
\end{abstract}

\pacs{64.60.Cn,05.70.Ln,82.20.-w,89.75.Da}
\maketitle

Even though it is widely believed that far from equilibrium
phase transitions can be classified, recent numerical results
reveal that we are still far from a satisfactory understanding of
even the relevant ingredients deciding to which class a given 
transition belongs. 
This is in particular true for transitions from a fluctuating phase to
one or several absorbing states (APT, for absorbing phase transitions)
where, in spite of a wealth of analytical and numerical studies,
rather little is known beyond the existence of the prominent
universality class of directed percolation (DP) 
\cite{HH-REVIEW}. In terms of 
reaction-diffusion processes, the DP class is often represented
by the simplest branching-annihilating reactions: $A\to 2A$, $A\to \emptyset$.

Following an early suggestion by Grassberger \cite{GRASS82}, a lot of attention
has been devoted recently to the case of binary or pair reaction-diffusion
processes  such as $2A\to (2+k)A$, $2A\to \emptyset$ or $A$, 
which involve two particles both for 
branching and annihilating 
\cite{HT97,CHS01,HH01,ODOR00,HH01bis,ODOR01bis,OMS01,ODOR02,MHHH02,NP02,DICK02,MUNOZ,HH-NEW}. 
Usually designated under the name of 
``pair contact process with diffusion'' (PCPD), it is still unclear
as of now if systems of this type can exhibit universal scaling laws
or even if (simple) scaling occurs at all at the APT. 
The origin of the strong
deviations to scaling observed is debated \cite{DICK02,HH-NEW}, 
and various conclusions
have been drawn ranging from
diffusion-rate dependent sub-classes \cite{ODOR00} 
to continuously-varying exponents \cite{NP02} to
no scaling to slow crossover to the DP class \cite{HH-NEW}. 
At the analytical level, a bosonic field theory of PCPD processes exists but
is not renormalizable whereas no fermionic version is available 
\cite{HT97,CHS01,MUNOZ}.
Completing this unsatisfactory picture, similar results were reported
recently for the ``triplet contact process with diffusion'' (TCPD)
$3A\to 4A$, $3A\to 2A$, where three particles are involved for both
reactions \cite{TCPD}.
Finally, the role of the conservation of parity of the number of particles
in APT is not well understood either: it is known to be relevant for some
one-dimensional DP-like processes, 
giving rise to the so-called ``parity-conserving'' (PC) class
\cite{PC,CT}, whereas
it was argued not to influence the critical properties of PCPD systems
\cite{PHK01}.

In this Letter, we report on a systematic numerical investigation of 
the reaction-diffusion processes $mA\to (m+k)A$, $nA \to (n-l)A$ 
in one space dimension,
from which we draw a considerably clarified picture of the above situation.
This is achieved thanks to a new class of very simple models which,
contrary to all the works cited above, abandon the ``fermionic constraint''
usually considered for both theoretical  and practical reasons
(to avoid the divergence of the density of particles 
in the active phase, and ---hopefully--- increase numerical
efficiency). Our results include the first evidence of 
universal (ordinary) scaling for PCPD and TCPD processes, the classification
of ``hybrid'' rules (i.e. those for which $m\ne n$), and further
insights into the conditions under which the conservation of the 
number of particles modulo 2 or 3 is able to change the ``reference''
scaling laws.

Following Hinrichsen \cite{TCPD}, 
we may write, for the $mA\to (m+k)A$, $nA \to (n-l)A$ 
processes, the following ``mean-field 
Langevin equation'' expected to govern the coarse-grained local density
$\rho$:
\begin{equation}
\partial_t\rho = a \rho^m - b \rho^n - c \rho^{m+1} + D \nabla^2\rho 
+\zeta({\bf x},t) \;,
\label{eq-mf}
\end{equation}
where $a$, $b$, and $c$ are positive constants related to the reaction rates
and $D$ is the diffusion constant. Note that of the two negative terms
in Eq.(\ref{eq-mf}) only the lowest-order one is relevant. 
As usual, the stochastic nature of the process is embodied in $\zeta$, 
a delta-correlated 
noise whose amplitude is a power of $\rho$ (thus ensuring the absence
of fluctuations in the $\rho=0$ absorbing phase):
\begin{equation}
\langle\zeta({\bf x},t)\zeta({\bf x}',t')\rangle = 
\Gamma \rho^\mu({\bf x},t)\delta^d({\bf x}-{\bf x}')\delta(t-t') \;.
\label{eq-noise}
\end{equation}
In the absence of branching (i.e. in the inactive phase leading to an
absorbing state), the noise dimension $\mu=n$ and the upper critical
dimension of the annihilation process is $d_{\rm c}=2/(n-1)$. 
In the critical region,
fluctuations are expected to increase the noise strength and thus to reduce
$\mu$ and increase $d_{\rm c}$. 
In the following, we restrict ourselves to $m,n\le4$
(in one dimension, non-trivial APT are mostly expected to occur if $n<4$).

\begin{figure}
\includegraphics[width=7.1cm]{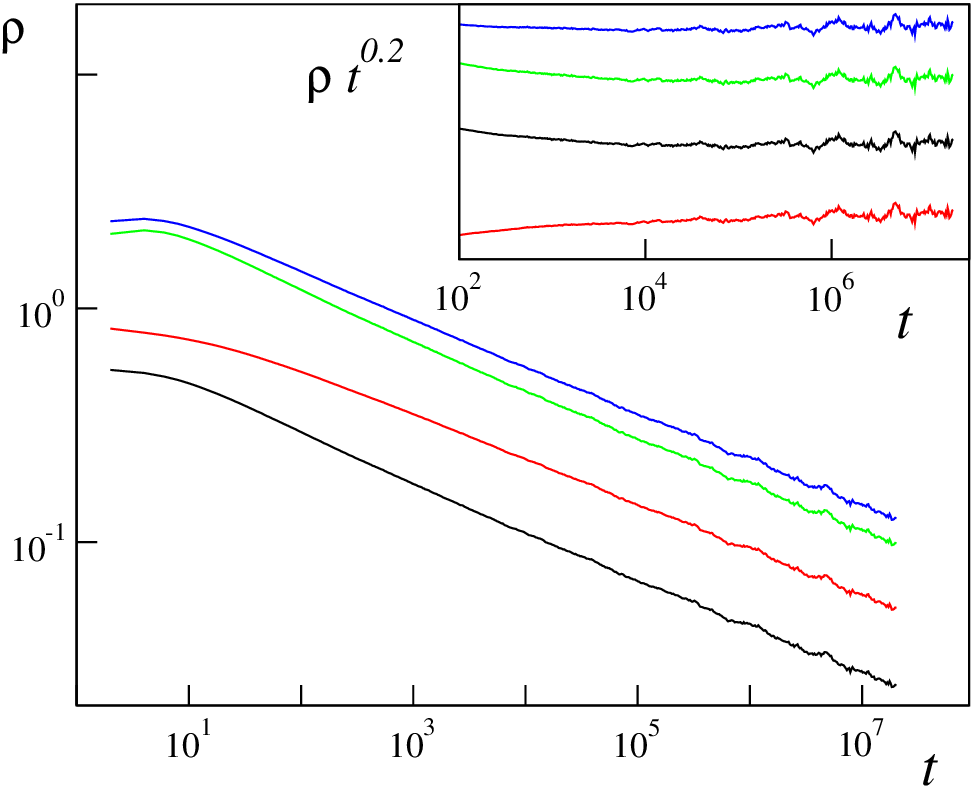}
\caption{Time decay of the order parameters at criticality for
the PCPD rule {\sf pp}12 ($p_{\rm c}=0.795410[5]$) in logarithmic scales. 
Single run on a system of $2^{22}$ sites, starting from 2 particles
on each site (other initial conditions do not influence the
asymptotic behavior). Different order parameters, from top to bottom:
density of particles, 
density of pairs (e.g., by summing all
particles of sites occupied by at least 2),
fraction of occupied sites, 
and fraction of sites occupied by at least a pair.
Inset: same, but for order parameters multiplied by $t^{0.2}$.}
\label{f1}
\end{figure}

Each of the branching-annihilating processes  $mA\to (m+k)A$, $nA \to (n-l)A$
is defined by the 4 integers $(m,n,k,l)$. Obviously, one must have
$0<n-l<m$ to insure the existence of at least one absorbing state.
For legibility, $m$ and $n$  will be coded below
by the letters {\sf s}, {\sf p}, {\sf t}, {\sf q} 
(for singleton, pair, triplet, quadruplet). 
The PCPD rule $2A\to 3A$, $2A\to \emptyset$ is thus noted {\sf pp}12,
and {\sf tt}$xx$ denote TCPD processes. 
The fermionic constraint adopted in most PCPD and TCPD models
studied so far can be seen as counter-productive: the actual implementation
of their Monte-Carlo simulations often lead to complicated rules,
inefficient for both code elaboration and simulation.
More importantly, the fermionic constraint may well be at the origin
of the strong deviations to scaling observed. 
Our branching-annihilating random walk (BARW) models 
are designed to bypass both of these problems. 
Particles of a single species $A$
evolve in parallel in two synchronized sub-steps: 
random walk on the lattice (diffusion) followed by on-site reaction.
For simplicity, in the following, the diffusion constant
is kept constant: all particles always jump to a randomly chosen
nearest-neighbor.
Let us consider, to describe our on-site reaction scheme, 
the PCPD rule {\sf pp}12. (The
generalization to all other rules studied here is straightforward.) 
Suppose that $n_A$ particles are present at a given site.
If $n_A=1$, nothing
happens (this is the main PCPD constraint). For $n_A\ge2$, each of
the $\lfloor n_A/2 \rfloor$ pairs into which the local population can be divided
branches with probability $p^{\lfloor n_A/2 \rfloor}$ (thus creating each time 
one new particle for this particular example) otherwise it annihilates.
The only parameter is $p$: for large $p$, branching is likely,
and one expects to be in the active phase. For small $p$, annihilation
dominates (indeed this is the only process at play for $p=0$), leading 
to an absorbing state. One key feature is the nonlinearity introduced
by raising $p$ to the power $\lfloor n_A/2 \rfloor$:
branching/annihilation is inhibited/enhanced  for large local populations,
preventing the divergence of population of usual bosonic models.
All the results presented below were obtained for reaction schemes of
this type, but we have checked that the functional form of the nonlinearity
as well as other details do {\it not} change the critical properties
observed.

As recommended when studying APT numerically, we first investigate,
for a given rule, the decay of the order parameter from some
highly active, correlation-less, initial condition in a large system. 
Above threshold, activity eventually reaches a constant level in time. 
Well below threshold,
one observes a decay typical of the annihilation process (exponential
for $n=1$, algebraic for $n>1$). At $p=p_{\rm c}$, in the usual framework,
one expects a non-trivial algebraic decay characterized by the
scaling exponent $\delta=\beta/\nu_{\parallel}$.
Once the threshold determined, other scaling exponents can
be estimated. In this work, we also give ---without showing the data---
our estimates of $z$,
obtained by the finite-size scaling of the mean lifetime of the system at 
threshold ($\langle\tau\rangle \sim L^z$), and of $\beta$, given by the
decay of the stationary density with the distance to threshold 
($\lim_{t\to\infty}\rho \sim (p-p_{\rm c})^\beta$).
More precise estimates will appear elsewhere
\cite{TBP}. In the following, the role of
conservation (modulo 2 or 3) of the number of particles is 
discussed at the end.
Before proceeding to higher-order rules, we report that {\sf ss}$xx$
rules are easily verified to exhibit DP-class scaling laws (not shown).

\begin{figure}
\includegraphics[width=7.1cm]{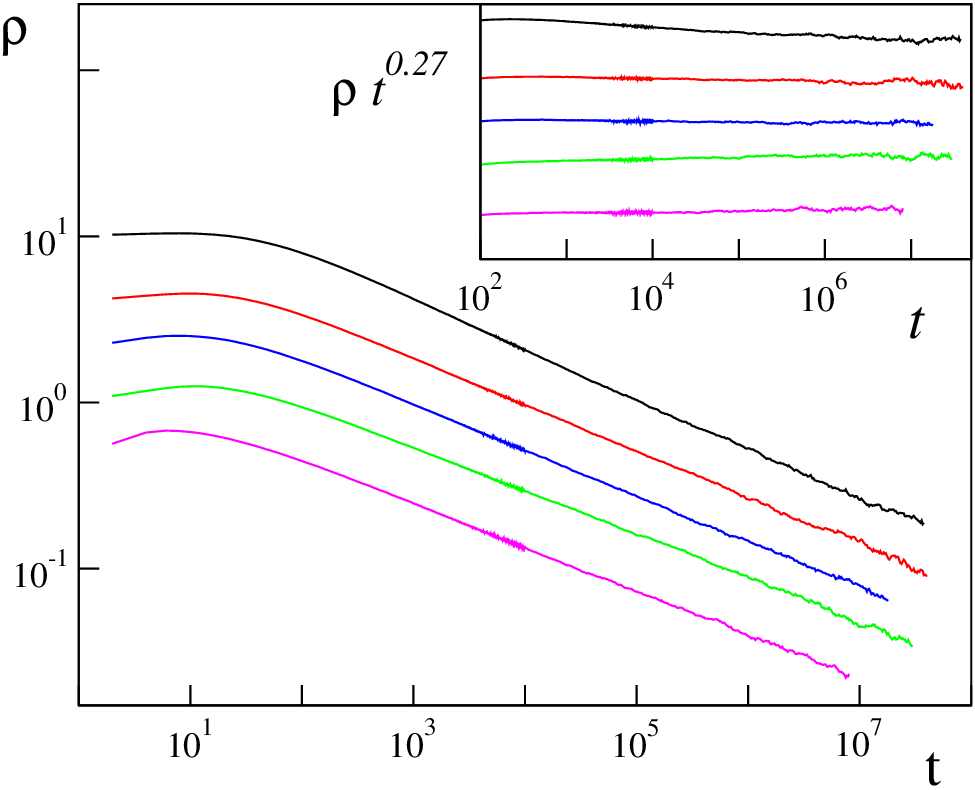}
\caption{Time decay of the total density of particles at criticality for
various TCPD rules in logarithmic scales. 
Single runs on a system of $2^{22}$ sites, starting from 3 particles
on each site.
From top to bottom: 
rule {\sf tt}11, $p_{\rm c}=0.53948(1)$;
rule {\sf tt}12, $p_{\rm c}=0.72899(1)$;
rule {\sf tt}13, $p_{\rm c}=0.822055(3)$;
rule {\sf tt}22, $p_{\rm c}=0.61900(2)$;
rule {\sf tt}33, $p_{\rm c}=0.70323(2)$;
Inset: same, but multiplied by $t^{0.27}$.
All data have been shifted for clarity.
}
\label{f2}
\end{figure}

In Fig.~\ref{f1}, we show the critical behavior 
of the PCPD rule {\sf pp}12. Quite easily, a clean algebraic decay in time
of all order parameters is observed over many decades with the
exponent $\delta^{\rm PCPD}=0.200(5)$.
At the determined threshold, the mean lifetime scales nicely
with system size, yielding the estimate $z^{\rm PCPD} = 1.70(4)$. 
Finally, the decay of the stationary order parameter with distance
to threshold allows to estimate $\beta^{\rm PCPD} = 0.375(10)$.
These exponent values appear to be universal: this has been checked
to the above numerical accuracy for rule {\sf pp}11 and with less 
care for several other rules.

Similar results were obtained for TCPD rules. Although scaling usually
sets in later than for the PCPD rules, it is well established over
a large range of scales. Again, scaling laws are found to be universal
within numerical accuracy. Fig.~\ref{f2} shows typical results obtained
for the decay of the particle density at threshold for five different
rules, from which we
estimate $\delta^{\rm TCPD}=0.27(1)$.
Other exponents are reported in Table~\ref{tab-exp}.

\begin{figure}
\includegraphics[width=7.1cm]{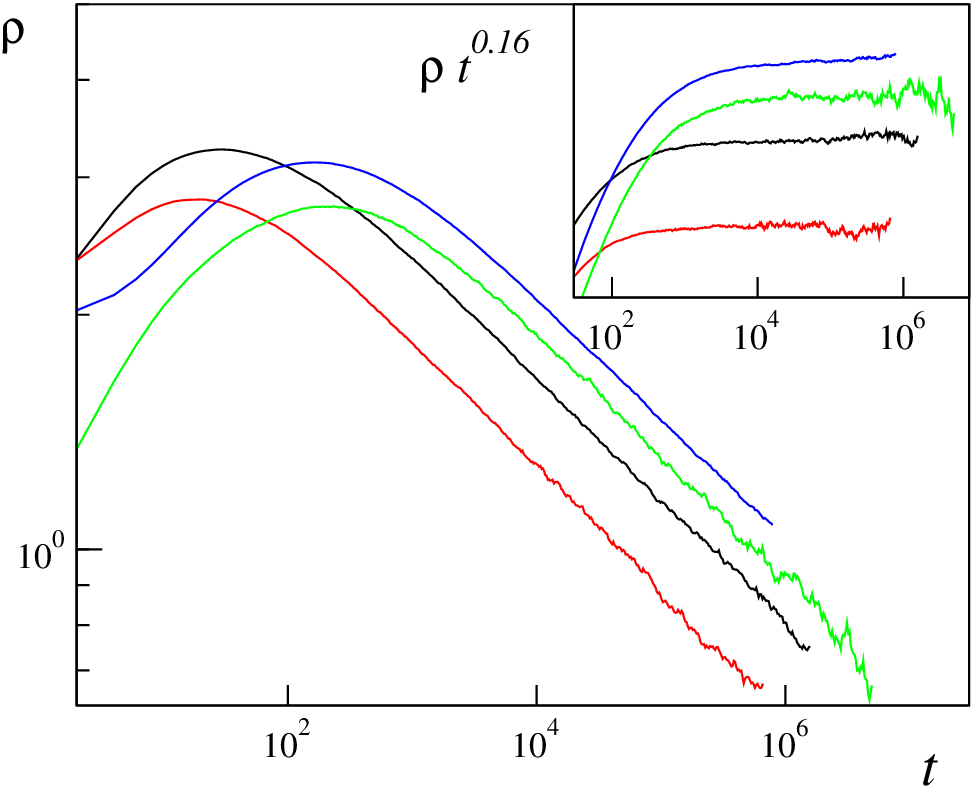}
\caption{Same as Fig.~\protect\ref{f2} but for hybrid rules $m>n$.
From top to bottom: 
rule {\sf tp}12, $p_{\rm c}=0.81156(1)$, $2^{24}$ sites;
rule {\sf ts}21, $p_{\rm c}=0.62121(2)$, $2^{20}$ sites;
rule {\sf tp}22, $p_{\rm c}=0.756625(5)$, $2^{22}$ sites;
rule {\sf ps}11, $p_{\rm c}=0.71529(1)$,  $2^{20}$ sites.
Inset: same, but multiplied by $t^{0.16}$.}
\label{f3}
\end{figure}

\begin{table}
\caption{\label{tab-exp}Critical exponents for basic classes.}
\begin{ruledtabular}
\begin{tabular}{llll}
Class & $\delta$ & $z$ & $\beta$\\ \hline
DP (from \protect\cite{JENSEN1}) & 0.1595 & 1.58 & 0.2765\\
PCPD & 0.200(5) & 1.70(5) & 0.37(2) \\
TCPD & 0.27(1) & 1.8(1) & 0.90(5)\\
PC (from \protect\cite{JENSEN2}) & 0.286 & 1.76 & 0.92\\
\end{tabular}
\end{ruledtabular}
\end{table}

Next, we consider hybrid rules for which $m>n$. At the mean-field level,
they are expected to present first-order transitions (see Eq.(\ref{eq-mf})).
Nevertheless, in one space dimension, all the cases we considered
show DP scaling, at least for the decay exponent $\delta$.
This is true for $n=1$ (simple radioactive decay), a case where 
we have studied rules {\sf ps}11, {\sf ts}21, and {\sf qs}11.
But this is also true for higher-order annihilation processes,
as shown by pair-annihilation rules {\sf tp}12 (2 variants) and {\sf qp}12,
as well as by triplet-annihilation rule {\sf qt}12.
As seen in Fig.~\ref{f3}, for high-order branching process
scaling sets in rather late, probably because the initial conditions
chosen are not ``optimal''.
That a given process for which a first-order APT is predicted at
mean-field level exhibits a continuous transition in low dimensions
is by no means surprising, especially in one dimension, where this has
been noticed early \cite{BBC}. Nevertheless, to our knowledge,
satisfactory analytical arguments for our finding are not available.

\begin{figure}
\includegraphics[width=7.1cm]{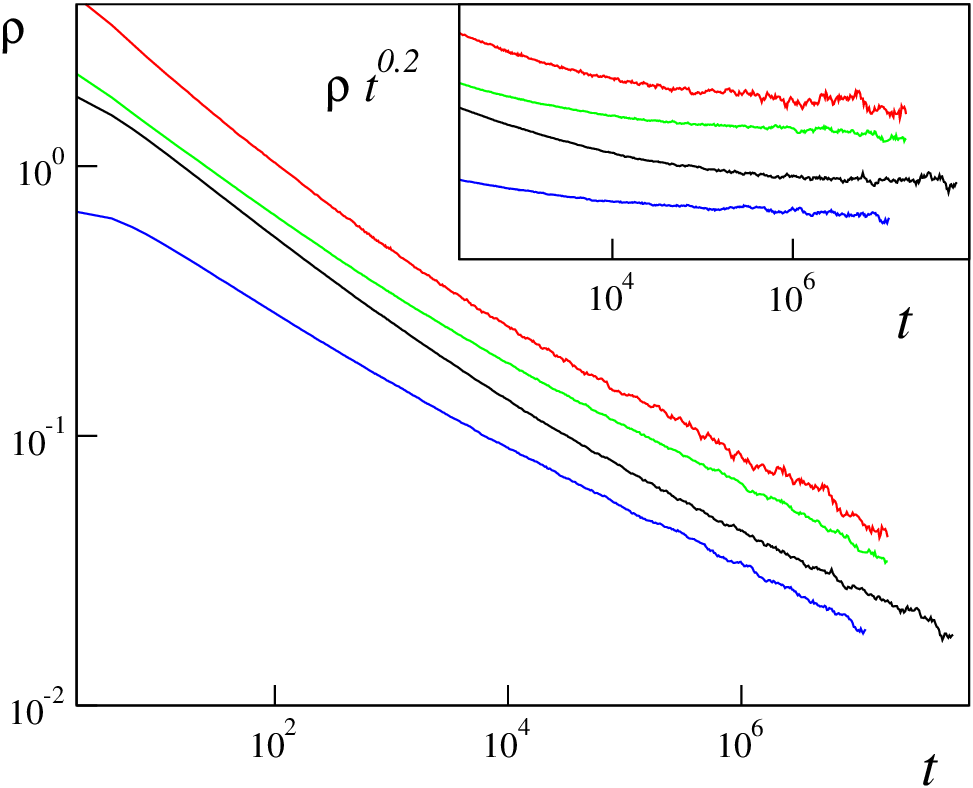}
\caption{Same as Fig.~\protect\ref{f2} but for {\sf pt}$xx$ hybrid rules.
From top to bottom: 
rule {\sf pt}42, $p_{\rm c}=0.56775(4)$;
rule {\sf pt}22, $p_{\rm c}=0.54640(2)$;
rule {\sf pt}12, $p_{\rm c}=0.33250(3)$;
rule {\sf pt}13, $p_{\rm c}=0.66210(2)$.
Inset: same, but multiplied by $t^{0.2}$.}
\label{f4}
\end{figure}

Hybrid rules for which the branching process is of lower order than
the annihilation reaction ($m<n$) are also easily investigated 
within our family of models. First, we note that whenever $n=4$,
no non-trivial APT was found in any of the
rules {\sf sq}$xx$, {\sf pq}$xx$, {\sf tq}$xx$ or {\sf qq}$xx$
that we have considered. In other words, $p_{\rm c}=0$ and the decay
is then that of the mean-field prediction ($t^{-1/3}$),
in agreement with the fact that the upper critical dimension of the 
decay process is $\frac{3}{4}<1$. For $m<n<4$, we find that
the universality class of the critical behavior is determined by
the order of the branching process: 
$m=1$ rules such as {\sf sp}12, {\sf st}13, {\sf st}23 
exhibit DP-scaling, in agreement with the analytical arguments of \cite{CT}.
Pair-branching rules ($m=2$)
{\sf pt}12 and {\sf pt}13, after a rather long crossover time,
show scaling compatible with the PCPD class (Fig.~\ref{f4}). 
That $m$ sets the universality class in this case is actually
not too surprising: it is clear
at the mean-field level of Eq.(\ref{eq-mf}), and can also be deduced from
``vertex generation arguments'' such as those developed in \cite{CT}.
 
We now pay particular attention to the
 rules for which the total number of 
particles is conserved modulo 2 (parity conservation) or modulo 3.
Phase space is then divided into 2 or 3 disjoint sectors,
each of which may or may not possess an absorbing state.
For example, the even sector of parity-conserving rule {\sf sp}22, 
the archetypical rule of the PC class,
includes an absorbing state (the empty configuration)
whereas its odd sector does not. 
For parity-conserving PCPD rule {\sf pp}22, on the other hand,
each sector has an absorbing state. 
We studied at least one rule in each of the relevant sub-families
(namely {\sf pp}, {\sf tt}, {\sf sp}, {\sf pt}, {\sf tp},
{\sf qp}, {\sf qt} for parity conservation, and
 {\sf st}, {\sf pt}, {\sf tt}, {\sf qt} for conservation $\mod 3$).
We find that whenever {\it every} sector includes an absorbing state,
mod2 or mod3 conservation does {\it not} change the ``reference''
scaling class. Thus rule {\sf pp}22 is in the PCPD class (as suggested
in \cite{PHK01}), rules {\sf tt}22 and {\sf tt}33 in the TCPD class
(Fig.~\ref{f2}).
Interestingly, rules {\sf tp}22, {\sf qp}22, {\sf qt}22, and {\sf qt}33
are found to be in the DP class, like their non-conserving counterparts,
and thus constitute the first examples of DP-class rules with mod2 
or mod3 conservation (Fig.~\ref{f3}).

Conversely, conserving rules with at least one
non-absorbing sector do not offer such a clear conclusion:
as expected, rules {\sf sp}22 and {\sf sp}42 fall into the PC class (not shown).
As argued in \cite{CT},
rule {\sf st}33 does not have a non-trivial APT. This is also observed
for rule {\sf pt}33, probably for similar theoretical
reasons. The only possible case left is that of parity-conserving
hybrid rules {\sf pt}$(2k)2$. There the even sector is non-absorbing
because the empty state ---which is absorbing--- cannot be reached from
any other configuration. Numerical simulations of 
rules {\sf pt}22 and {\sf pt}42, however, do {\it not} show any significant
difference from their non-conserving cousins (Fig.~\ref{f4}). This result,
which needs confirmation due to the lateness of scaling, seems to refute 
the suggestion of \cite{PHK01} that influence of parity conservation
may be equivalent to having a non-absorbing sector of phase space.
Instead we would like to propose that rules {\sf sp}$(2k)2$ are the only
ones in the PC class because they are the only BARW processes equivalent
to generalized two-state Voter models, as defined in \cite{VOTPRL}, i.e. 
spin models with up/down symmetry, no bulk noise, and
an order/disorder transition. This remark, which will be developed in
\cite{TBP2}, calls for renaming the PC class the Voter class.

\begin{table}
\caption{\label{tab-class}Universality class of BARW processes.
``$\emptyset$'' stands for no non-trivial APT ($p_{\rm c}=0$).
The second markings indicate the change of class whenever 
mod2 or mod3 conservation plays a role. See text for details.}
\begin{ruledtabular}
\begin{tabular}{lllll}
m$\diagdown$n & 1 & 2 & 3 & 4\\
\hline
1 & {\bf DP} & DP$\diagup$PC & DP$\diagup\emptyset$  & $\emptyset$\\
2 & DP       & {\bf PCPD}    & PCPD$\diagup\emptyset$ & $\emptyset$\\
3 & DP       & DP            & {\bf TCPD} & $\emptyset$\\
4 & DP       & DP            & DP         & $\emptyset$\\
\end{tabular}
\end{ruledtabular}
\end{table}

We now summarize and discuss our results. 
We first note that our BARW models, at criticality, 
exhibit ordinary scaling after some crossover scale,
and do not seem to be plagued by the strong deviations
observed with even the simplest fermionic PCPD or TCPD rules
\cite{DICK02,HH-NEW}.
Numerical results of basic fermionic models can be shown to 
converge to the critical behavior found here \cite{TBP}.
To our numerical accuracy, the scaling exponents recorded at criticality
lead to conclude to the existence of three basic universality classes:
DP, PCPD, and TCPD. Table~\ref{tab-exp} summarizes our current estimates
of the basic exponents $\delta$, $z$, and $\beta$, 
pending more precise ones \cite{TBP}.
Extrapolating from our numerical findings, we believe that
the critical behavior of all processes considered here is as follows
(Table~\ref{tab-class}):
rules with four-particle annihilation ($n=4$) do not have
non-trivial APT, indicating (in partial agreement with \cite{TCPD})
that $d_{\rm c}\le1$ in this case. 
Hybrid processes fall into one of these basic classes:
$m>n$ rules exhibit DP critical behavior, whereas the class of $m<n$ rules
is set by $m$. Finally, we have shown that mod2 (parity) or mod3 conservation
seems to act only on {\sf sp}$(2k)2$ rules, suggesting that the PC class
should be seen as the (one-dimensional, generalized) Voter class.

Obviously, our numerical findings need to be confirmed by analytical 
approaches. We hope that, at the very least, they will trigger
some new lines of attack to the difficult issues at play.
At the numerical level, ongoing work aims at obtaining more comprehensive
results, including the study of spreading exponents and the case
of higher space dimensions. 

We thank F. van Wijland for useful discussions.

\end{document}